\documentclass[aps,pre,twocolumn,groupedaddress,amsmath,showpacs,amssymb]{revtex4}
    \usepackage{dcolumn}
    \usepackage{bm}
    \usepackage{graphicx}
    \usepackage{dcolumn}
\newcommand\beq{\begin{equation}}
\newcommand\eeq{\end{equation}}
\newcommand\beqa{\begin{eqnarray}}
\newcommand\eeqa{\end{eqnarray}}

\newcommand{\gd}{\dot{\gamma}}

\newcommand{\wa}{\mu}

\newcommand{\wz}{\epsilon}

\begin{document}


\title{Does the Chapman--Enskog expansion for sheared granular gases
converge?}


\author{Andr\'es Santos}
\email{andres@unex.es} \homepage{http://www.unex.es/fisteor/andres/}
\affiliation{Departamento de F\'{\i}sica, Universidad de
Extremadura, E-06071 Badajoz, Spain}



\date{\today}

\begin{abstract}
The fundamental question addressed in this paper is whether the
partial Chapman--Enskog expansion $P_{xy}=-\sum_{k=0}^\infty \eta_k
\left({\partial u_x}/{\partial y}\right)^{2k+1}$ of the shear stress
converges or not for a gas of inelastic hard spheres. By using a
simple kinetic model it is shown that, in contrast to the elastic
case, the above series does converge, the radius of convergence
increasing with inelasticity. It is argued that this paradoxical
conclusion is not an artifact of the kinetic model and can be
understood in terms of the time evolution of the scaled shear rate
in the uniform shear flow.
\end{abstract}

\pacs{45.70.Mg, 
 47.50.-d, 
 05.20.Dd 
 51.10.+y} 


\maketitle


The hydrodynamic description of conventional fluids is usually
restricted to the Navier--Stokes (NS) constitutive equations
\cite{dGM84}. For instance, if the flow is incompressible (i.e.,
$\nabla\cdot\mathbf{u}=0$, where $\mathbf{u}$ is the flow velocity),
Newton's law establishes a linear relationship $P_{xy}=-\eta_0
{\partial u_x}/{\partial y}$ between the shear stress $P_{xy}$ and
the shear rate $\partial u_x/\partial y$, where $\eta_0$ is the
shear viscosity and it has been assumed that $\partial u_y/\partial
x=0$. The NS constitutive equations  represent excellent
approximations in most of the physical situations of experimental
interest, even if the regime is turbulent \cite{D89}. In the case of
a dilute gas, they can be justified under the assumption that the
smallest  of the characteristic hydrodynamic lengths ($L$)
associated with the gradients of density ($n$), temperature ($T$),
and flow velocity ($\mathbf{u}$) is much larger than the mean free
path $\ell$ of the gas particles, i.e., $\wa\equiv \ell/ L\ll 1$. As
a matter of fact, the Chapman--Enskog (CE) method provides a
systematic scheme to obtain the \emph{normal} solution of the
Boltzmann equation as an expansion in powers of the uniformity
parameter (or Knudsen number) $\wa$ \cite{CC70}. The leading terms
in the CE expansion yield the NS constitutive equations, with the
bonus of providing expressions for the transport coefficients (like
the shear viscosity $\eta_0$) in terms of the microscopic properties
and of the hydrodynamic quantities.

A fundamental question concerning the CE method is the nature
(convergent versus divergent) of the expansion in powers of $\wa$.
More than 40 years ago Grad \cite{G63} provided compelling arguments
on the asymptotic character of the CE expansion. This means that, as
expected on physical grounds, the CE series truncated at a given
order (e.g., NS, Burnett, super-Burnett, \ldots) becomes closer and
closer to the true value as $\wa$ becomes smaller and smaller. Of
course, the asymptotic character of the CE expansion does not imply
(but is implied by) the stronger condition of convergence. However,
McLennan \cite{M65} was able to prove convergence of the
\emph{partial} sum of the CE series made of \emph{linear} terms
(i.e., terms of the form $\nabla_{i_1}\nabla_{i_2}\ldots
\nabla_{i_k}A$, where $A=n$, $T$, or $\mathbf{u}$) for a general
class of cutoff potentials.

What about the \emph{nonlinear} terms of the  series? To be more
specific, let us consider the following subclass of the full CE
series of the shear stress:
\beq
P_{xy}=-\sum_{k=0}^\infty \eta_k \left({\partial u_x}/{\partial
y}\right)^{2k+1},
\label{2}
\eeq
where $\eta_0$ is the NS shear viscosity, $\eta_1$ is a
super-Burnett coefficient, and so on. The full CE series of $P_{xy}$
reduces to the partial series  \eqref{2} if (and only if) $\partial
u_i/\partial x_j=\gd \delta_{ix}\delta_{jy}$ and $\nabla n=\nabla
T=\nabla\gd=\mathbf{0}$, i.e., the only nonzero hydrodynamic
gradient is a uniform shear rate $\gd\equiv \partial u_x/\partial
y$. Interestingly enough, there exists a physical state (the
so-called simple or uniform shear flow, USF) that is consistent with
those conditions \cite{LE72,GS03}. In that state, the density and
the shear rate are constant in time but the temperature increases
due to  viscous heating. The identification of the characteristic
hydrodynamic length $L$ is (except for a numerical factor)
unambiguous: $L\sim \sqrt{2T/m}/\gd$ (where $m$ is the mass of a
particle), so that the uniformity parameter becomes $\wa=\gd/\nu$,
where $\nu\sim \sqrt{2T/m}/\ell$ is a characteristic collision
frequency. Thus, the CE expansion \eqref{2} can be recast into the
dimensionless form
\beq
P_{xy}/nT=- \wa F(\wa^2),\quad F(z)=\sum_{k=0}^\infty c_k z^k,
\label{13}
\eeq
where $c_k\equiv (\eta_k/nT)\nu^{2k+1}$. Despite its simple
definition, the USF state is complex enough to prevent an exact
solution of the nonlinear Boltzmann equation. However, the problem
becomes solvable in the framework of the Bhatnagar--Gross--Krook
(BGK) model kinetic equation \cite{GS03} and the solution shows
that, for a wide class of repulsive potentials, the CE expansion
\eqref{2} is \emph{divergent} \cite{SBD86}. More specifically, the
dimensionless coefficients $c_k$ behave for large $k$ as $|c_k|\sim
(2/d)^k k!$ in the case of $d$-dimensional hard spheres.

In the preceding paragraphs we have considered conventional gases
made of particles that collide elastically. On the other hand, the
same issues discussed above can be applied to granular gases, i.e.,
large assemblies of (mesoscopic or macroscopic) particles which
collide inelastically and are maintained in fluidized states. Apart
from their practical interest, granular gases are  physical systems
worth studying at a fundamental level because they are intrinsically
out of equilibrium and exhibit a wide repertoire of complex and
exotic behavior \cite{K00,K99}. Nonequilibrium
statistical-mechanical concepts and tools, in particular the kinetic
theory approach based on the Boltzmann and Enskog equations suitably
modified to account for inelastic collisions, have proven to be
useful to understand the behavior of granular gases \cite{BP04}.
Most of the theoretical efforts have focused on a minimal model of
granular gases consisting of smooth inelastic hard spheres
characterized by a constant coefficient of normal restitution
$\alpha<1$. Specifically, the CE method has been applied to the
(inelastic) Boltzmann and Enskog kinetic equations and the NS
transport coefficients have been derived \cite{BDKS98}.

The question I want to address in this Letter is, does the nonlinear
subclass  of the full CE expansion  converge in the case of a gas of
inelastic hard spheres? Given that the answer is negative when the
gas is made of elastic hard spheres ($\alpha=1$) \cite{SBD86}, the
strong challenges to the validity of hydrodynamics in granular media
\cite{K99}, and the inherently non-Newtonian nature of the steady
USF of granular gases \cite{SGD04}, it seems plausible to expect
that the (partial) CE series \eqref{2} is divergent for granular
gases. It will be shown below that, paradoxically, the series
\eqref{2} does converge in the case of inelastic hard spheres and
that, in fact, the radius of convergence increases with increasing
inelasticity.

The energy balance equation for inelastic hard spheres in the USF is
\begin{equation}
\partial _{t}T(t)=-({2}/{dn})\gd P_{xy}(t) -\zeta(t) T(t),
\label{2.8}
\end{equation}
where $\zeta$ is the so-called cooling rate \cite{BP04}. {}From the
Boltzmann equation it is possible to show that it is approximately
given by $\zeta=\frac{d+2}{4d}(1-\alpha^2)\nu_0$,  $\nu_0$ being an
effective collision frequency for elastic spheres. The cooling term
on the right-hand side of Eq.\ \eqref{2.8} competes with the viscous
heating term, so that, depending on the initial state and the value
of $\gd$, the temperature either grows or decreases with time until
a steady state is eventually reached \cite{SGD04,AS07}. In order to
relate the shear stress $P_{xy}(t)$ to the shear rate $\gd$ and to
the temperature $T(t)$, and thus analyze the series representation
\eqref{13}, let us replace the Boltzmann equation by the BGK-like
kinetic model \cite{BDS99}
\beq
(\partial_t
+\mathbf{v}\cdot\nabla)f=-\nu(f-f_0)+({\zeta}/{2})\partial_\mathbf{v}\cdot(\mathbf{V}f),
\label{n1}
\eeq
where $f$ is the velocity distribution function, $\mathbf{V}\equiv
\mathbf{v}-\mathbf{u}$ is the peculiar velocity, $f_0$  is the local
version of the homogeneous cooling state distribution {\cite{BP04}}
(parameterized by the actual fields $n$, $\mathbf{u}$, and $T$), and
$\nu$ is an effective collision frequency. {Here the simple choice
$\nu=\frac{1+\alpha }{2}\nu_0$ is made, so that $\wz\equiv
\zeta/\nu=\frac{d+2}{2d}(1-\alpha)$.} Taking moments in Eq.\
\eqref{n1} one gets
\beq
\begin{array}{ll}
\partial_t P_{xy}(t)=&-\gd
P_{yy}(t)-[\nu(t)+\zeta(t)]P_{xy}(t),\\
\partial_t
P_{yy}(t)=&n\nu(t)T(t)-[\nu(t)+\zeta(t)]P_{yy}(t),
\end{array}
\label{6}
\eeq
where $P_{yy}$ is a normal stress. {Note that the explicit form of
$f_0$ is not needed in the derivation of   Eq.\ \protect\eqref{6}
and so no assumption of $f$ or $f_0$ being close to a Maxwellian is
taken. Equation \protect\eqref{6}, with $\nu=\frac{1+\alpha
}{2}\left[ 1-\frac{d-1}{2d} (1-\alpha )\right]\nu_0$, is also
obtained from the Boltzmann equation in Grad's moment approximation
\cite{SGD04}.}
\begin{table*}
\caption{\label{table1}CE coefficients $c_k$ for $0\leq k\leq 6$,
steady-state value of the square uniformity parameter, $z_s$, and
estimates of the parameters $a$ and $\ln A$ associated with the
asymptotic behavior of $|c_k|$.}
\begin{ruledtabular}
\begin{tabular}{ccccccccccc}
$\alpha$& $c_0$& $c_1$& $c_2$& $c_3$& $c_4$& $c_5$& $c_6$&$z_s$&$a$&$\ln A$\\
\hline
$0.5$&$0.827586$&$-0.380155$&$0.0677085$&$0.0404196$&$-0.00251705$&$-0.0119767$&$-0.00399593$&$1.25434$&$15.8$&$60.3$\\
$0.7$&$0.888889$&$-0.536326$&$0.146688$&$0.110743$&$-0.0180970$&$-0.0727681$&$-0.0336082$&$0.585938$&$8.21$&$7.08$\\
$0.9$&$0.960000$&$-0.798195$&$0.377471$&$0.376385$&$-0.188551$&$-0.778076$&$-0.618849$&$0.146701$&$15.8$&$22.6$\\
$0.99$&$0.995851$&$-0.976777$&$0.627438$&$0.720350$&$-0.694311$&$-3.20037$&$-3.75620$&$0.0127092$&$95.9$&$283$\\
$1$&$1$&$-1$&$0.666667$&$0.777778$&$-0.814815$&$-3.82716$&$-4.72428$&$0$&---&---\\
\end{tabular}
\end{ruledtabular}
\end{table*}
Equations \eqref{2.8} and \eqref{6} constitute a closed set of
equations for the evolution of $T$, $P_{xy}$, and $P_{yy}$. In order
to describe the hydrodynamic regime  and analyze the CE expansion we
must focus on the nonlinear dependence of the {scaled viscosity
function $F(\wa^2(t))\equiv -P_{xy}(t)/nT(t)\wa(t)$ as a function of
the scaled shear rate (or uniformity parameter)
$\wa(t)\equiv\gd/\nu(t)$. Elimination of time in favor of $z\equiv
\wa^2\propto 1/T$  in Eqs.\ \eqref{2.8} and \eqref{6} yields the
following single second-order ordinary differential equation:}
\begin{widetext}
\beq
1-(1+\frac{2}{d}zF)^2F=\frac{1}{2}(\wz-\frac{2}{d}zF)\{z\partial_z
[(\wz-\frac{2}{d}zF) (F+2z\partial_zF)]+(1+\frac{2}{d}zF)
(F+2z\partial_zF)+2z\partial_z[(1+\frac{2}{d}zF)F]\}.
\label{12}
\eeq
\end{widetext}
 Insertion of the series expansion $F(z)=\sum_{k=0}^\infty c_k z^k$
into Eq.\ \eqref{12} allows one to get a recursive relation
expressing the coefficient $c_k$ for $k\geq 1$ in terms of the
lower-order coefficients $\{c_{k'}, 0\leq k'\leq k-1\}$. With the
help of a computer algebra system, one can obtain the coefficients
$c_k$,  for given $d$ and $\alpha$, {as exact rational numbers} up
to a maximum value of $k$ limited by computer time and internal
memory. The results of $c_k$ presented here have been obtained  for
$k\leq 400$.

\begin{figure}
\includegraphics[width=\columnwidth]{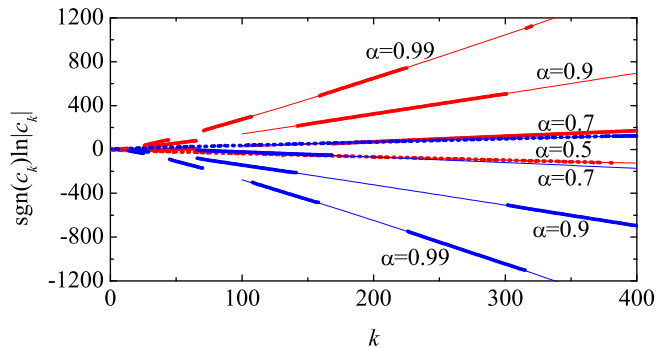}
\caption{(color online) Plot of $\text{sgn}(c_k)\ln|c_k|$ for $d=3$
and $\alpha=0.5$, $0.7$, $0.9$, and $0.99$. The thin solid lines
represent $\pm \left(-k\ln z_s-a\ln k+\ln A\right)$, where $a$ and
$\ln A$ are fitting parameters.
\label{fig1}}
\end{figure}
\begin{figure}
\includegraphics[width=\columnwidth]{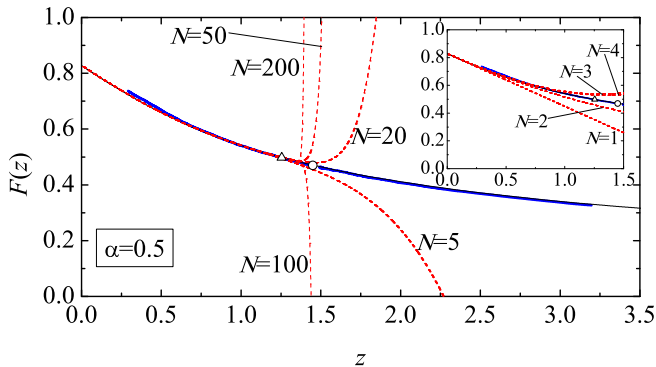}
\caption{(color online) Scaled viscosity function $F(z)$ for $d=3$
and $\alpha=0.5$. The thin solid line represents the numerical
solution of Eq.\ \eqref{12}, while the dashed lines represent the
truncated CE expansion $F_N(z)$ with $N=5$, $20$, $50$, $100$, and
$200$.  The thick solid line represents the results obtained from
Monte Carlo simulations of the Boltzmann equation
\protect\cite{AS07}. {The steady-state points $(z_s,F_s)$ obtained
from the kinetic model and  the simulations are indicated by the
triangle and the circle, respectively. The inset includes $F_N(z)$
with $N=1$--$4$ in the region $0\leq z\leq 1.5$}.
\label{fig2}}
\end{figure}
The first few coefficients $c_k$ are listed in Table \ref{table1}
for $d=3$ and $\alpha=0.5$, $0.7$, $0.9$, $0.99$, and $1$ (elastic
hard spheres). As is already apparent from Table \ref{table1}, the
magnitude of the coefficients increases as the system becomes less
inelastic, this effect being more and more dramatic with increasing
$k$. For instance, the value of $|c_{20}|$ is $1.2\times 10^{-5}$,
$5.2\times 10^{-3}$, $2.1\times 10^{5}$, $1.4\times 10^{11}$, and
$1.1\times 10^{12}$ for $\alpha=0.5$, $0.7$, $0.9$, $0.99$, and $1$,
respectively. Figure \ref{fig1} shows a log-normal plot of $|c_k|$
for the four inelastic systems considered. The results are
consistent with an asymptotic behavior of the form $|c_k|\approx A
k^{-a} z_s^{-k}$, where $z_s=\frac{d}{2}\wz(1+\wz)^2$ is the
steady-state value of the square uniformity parameter. The exponent
$a$ and the amplitude $A$ are estimated by a linear fit of a plot of
$\ln(|c_k|z_s^k)$ versus $\ln k$ for $100\leq k\leq 400$, the fit
being less robust in the case $\alpha=0.5$ than in the other cases.
The numerical values of $z_s$, as well as of the estimates of $a$
and $\ln A$ are also included in Table \ref{table1}. {In principle,
the viscosity function $F(z)$ is different for each dimensionality.
However, inspection of Eq.\ \eqref{12} shows that, for a given value
of $\wz$, the dependence of $F$ on $z$ and $d$ occurs only through
the  combination $z/d$. Therefore, seen as a function of $\wz$ and
$z/z_s$, $F(z)$ is independent of $d$}.

The large-$k$ asymptotic behavior $\ln|c_k|\sim -k\ln z_s$ for
$\alpha<1$ implies that the CE series \eqref{2}  is
\emph{convergent} for inelastic hard spheres. Moreover, the radius
of convergence is $\partial u_x/\partial y=\nu z_s^{1/2}$  and thus
increases with inelasticity. Equation \eqref{12} shows that $z=z_s$
is indeed a singular point since the steady state $(z_s,F_s)$, with
$F_s=(1+\wz)^{-2}$, is the solution of
$1-\left(1+\frac{2}{d}zF\right)^2F=\wz-\frac{2}{d}zF=0$. What is
relevant here is that $z=0$ is a regular point (except in the
elastic case) and $z=z_s$ is the singularity of $F(z)$ in the
complex $z$ plane closest to the origin. The convergent character of
the series \eqref{13} for the most inelastic case considered here
($\alpha=0.5$) is illustrated in Fig.\ \ref{fig2}, where the
numerical solution \cite{SGD04} of Eq.\ \eqref{12}  is compared with
the \emph{truncated} CE series $F_N(z)\equiv \sum_{k=0}^N c_k z^k$
for several values of the truncation order $N$. One can observe that
the truncated series agree among themselves and with the numerical
solution for $z<z_s$. For $z>z_s$, however, the CE expansion becomes
useless and one must determine $F(z)$ numerically from Eq.\
\eqref{12}. An alternative method, successfully used in the elastic
case \cite{SBD86}, would consist of expanding $F(z)$ around the
point at infinity as $F(z)=z^{-2/3}\sum_{k=0}^\infty \overline{c}_k
z^{-k/3}$. Figure \ref{fig2} also includes results recently obtained
from Monte Carlo simulations of the Boltzmann equation \cite{AS07},
which show that the predictions of Eq.\ \eqref{12} are
quantitatively accurate, even for strong inelasticity, {except that
Eq.\ \eqref{12} underestimates the location of the steady-state
point $(z_s,F_s)$}. {The simulation curve shown in Fig.\
\protect\ref{fig2} is the collapse of data obtained by the direct
simulation  Monte Carlo   method starting from 20 different initial
conditions (ten for $z<z_s$ and ten for $z>z_s$), letting the system
evolve in time, and discarding the kinetic transients lasting a few
mean free times. This explains why the simulation curve does not
reach the point $z=0$. For simulation details the reader is referred
to Ref.\ \protect\cite{AS07}.}

Is the paradoxical regularization by inelasticity of the CE series
\eqref{2} an artifact of the kinetic model? The following physical
argument suggests that this is not the case. Let us assume that the
reference homogeneous state is perturbed at time $t=0$ by a
\emph{weak} shear rate $\gd$. {}Thus the homogeneous state ($\wa=0$)
is stable or unstable against this USF perturbation depending on
whether the time-dependent uniformity parameter $\wa(t)=\gd/\nu(t)$
asymptotically goes to zero or grows with time. The first situation
occurs in a conventional gas of elastic hard spheres since the
viscous heating produces a monotonic increase of temperature [cf.\
Eq.\ \eqref{2.8} with $\zeta=0$] and thus $\wa(t)\to 0$. However, in
a gas of inelastic hard spheres it is always possible that the
perturbation is small enough to make the cooling rate prevail over
the viscous heating rate  and so the temperature keeps decreasing in
time [cf.\ Eq.\ \eqref{2.8} with $\zeta>0$] until the steady state
is eventually reached; thus the uniformity parameter grows, i.e.,
$\wa(t)\to z_s^{1/2}$. Since the CE expansion \eqref{13} measures
the departure from the reference homogeneous state ($\wa=0$), it is
reasonable to expect that the series diverges if $\wa(t)$ goes to
zero while it converges if $\wa(t)$ increases with time.

The above heuristic argument can also be applied to the simplest CE
series in a compressible flow, namely
\beq
P_{yy}-nT=-\sum_{k=0}^\infty \eta_k' \left({\partial u_y}/{\partial
y}\right)^{k+1}.
\label{14}
\eeq
The full CE expansion of $P_{yy}$ reduces to the series \eqref{14}
in the uniform longitudinal flow (ULF) characterized by \cite{GK96}
$\partial u_y/\partial y=\gd(t)=\gd_0/(1+\gd_0t)$ and
$n(t)=(n_0/\gd_0)\gd(t)$. The exact energy balance equation,
$\partial _{t}T(t)=-({2}/{dn})\gd(t) P_{yy}(t)-\zeta(t) T(t)$, shows
that the uniformity parameter $\wa(t)=\gd(t)/\nu(t)$ increases with
time if $\gd_0>0$, both for elastic and inelastic collisions.
However, if $\gd_0<0$, $|\wa(t)|\to 0$ in the elastic case whereas
$|\wa(t)|$ increases toward a stationary value in the inelastic
case. Thus, the CE expansion \eqref{14} is expected to diverge for
conventional gases and converge for granular gases. This expectation
is confirmed by an analysis of the ULF based again on the kinetic
model \eqref{n1}.

It is important to note that the dependence  on $n$, $T$, and
$\alpha$ of the transport coefficients $\eta_k$ and $\eta_k'$ in the
partial CE expansions \eqref{2} and \eqref{14}, respectively, is not
influenced by the specific state under consideration. Accordingly,
the series \eqref{2} and \eqref{14} converge or diverge irrespective
of whether the system is in the USF, the ULF, or in any other state.
The advantage of the USF and ULF is that the full CE series of the
shear stress and the normal stress reduce to the partial series
\eqref{2} and \eqref{14}, respectively, thus allowing us to explore
their character in a rather detailed way. {Regarding the CE
subseries made of linear terms, it is reasonable to expect that, as
in the elastic case, it also converges for inelastic collisions,
especially if one takes into account the exact mapping between the
inelastic and elastic versions of the Enskog--Lorentz equations
\cite{SD06}.}

The CE expansion is the main route to hydrodynamics and so its
convergence or divergence has a renewed interest in granular gases
in view of some debate on the applicability of hydrodynamics to this
class of nonequilibrium systems \cite{K99}. I expect that this
Letter can contribute to a clarification of this controversial issue
{by presenting a case study where the application of the CE
expansion to describe the nonlinear regime might have a larger
practical interest in granular  than in conventional gases}.

I am grateful to J. W. Dufty and V. Garz\'o for useful comments on
an early version of this paper. This work has been supported by the
Ministerio de Educaci\'on y Ciencia (Spain) through Grant No.\
FIS2007--60977 (partially financed by FEDER funds) and by the Junta
de Extremadura (Spain) through Grant No.\ GRU07046.


\begin{thebibliography}{99}

\bibitem{dGM84}
S. R. de Groot and P. Mazur, \emph{Nonequilibrium Thermodynamics}
(Dover, New York, 1984).

\bibitem{D89}
R. J. Donnelly, ``Fluid Dynamics,'' in \emph{A Physicist's Desk
Reference}, ed.\ by H. L. Anderson (American Institute of Physics,
New York, 1989), pp.\ 196--209.

\bibitem{CC70}
S. Chapman and T. G. Cowling, \emph{The Mathematical Theory of
Non-uniform Gases} (Cambridge University Press, Cambridge, 1970).

\bibitem{G63}
H. Grad, Phys. Fluids \textbf{6}, 147 (1963).

\bibitem{M65}
J. A. McLennan, Phys. Fluids \textbf{8}, 1580 (1965); G. Scharf,
Helv. Phys. Acta \textbf{40}, 929 (1967); \textbf{42}, 5 (1969).

\bibitem{LE72}
A. W. Lees and S. F. Edwards, J. Phys. C: Solid State Phys.
\textbf{5}, 1921 (1972).

\bibitem{GS03}
V. Garz\'o and A. Santos, \emph{Kinetic Theory of Gases in Shear
Flows. Nonlinear Transport} (Kluwer, Dordrecht, 2003), and
references therein.

\bibitem{SBD86}
A. Santos, J. J. Brey, and J. W. Dufty,  Phys. Rev. Lett.
\textbf{56}, 1571 (1986); A. Santos and J. J. Brey, Physica A
\textbf{174}, 355 (1991).

\bibitem{K00}
{H. M. Jaeger, S. R. Nagel, and R. Behringer}, {Phys. Today}
{\textbf{49}} (4), {32} ({1996}); {Rev. Mod. Phys.} {\textbf{68}},
{1259} ({1996});  {J. M. Ottino and  D. V. Khakhar}, {Annu. Rev.
Fluid Mech.} {\textbf{32}}, {55} ({2000}); {A. Kudrolli}, {Rep.
Prog. Phys.} {\textbf{67}}, {209} ({2004}).

\bibitem{K99}
{L. Kadanoff}, {Rev. Mod. Phys.} {\textbf{71}}, {435} ({1999}).

\bibitem{BP04}
{N. Brilliantov and  T. P\"oschel}, {\emph{Kinetic Theory of
Granular Gases}} ({Oxford University Press, Oxford}, {2004});
  {C. S. Campbell}, {Annu. Rev. Fluid Mech.} {\textbf{22}}, {57} ({1990});
{J. W. Dufty}, {J. Phys.: Condens. Matt.} {\textbf{12}} {A47}
({2000}); e-print arXiv:0709.0479; {I. Goldhirsch}, {Annu. Rev.
Fluid Mech.} {\textbf{35}}, {267} ({2003}).

\bibitem{BDKS98}
J. J. Brey, J. W. Dufty, C. S. Kim, and A. Santos,  Phys. Rev. E
\textbf{58}, 4638 (1998); V. Garz\'o and J. W. Dufty, Phys. Rev. E
\textbf{59}, 5895 (1999); Phys. Fluids \textbf{14}, 1476 (2002); V.
Garz\'o, J. W. Dufty, and C. M. Hrenya, Phys. Rev. E \textbf{76},
031303 (2007); V. Garz\'o, C. M. Hrenya, and J. W. Dufty, Phys. Rev.
E \textbf{76}, 031304 (2007); S. H. Noskowicz, O. Bar-Lev, D.
Serero, and I. Goldhirsch, Europhys. Lett. \textbf{79}, 60001
(2007).

\bibitem{SGD04}
A.  Santos, V. Garz\'o, and J. W. Dufty,   Phys. Rev. E \textbf{69},
061303 (2004).


\bibitem{AS07}
A. Astillero and A. Santos,  Europhys. Lett. \textbf{78}, 24002
(2007).

\bibitem{BDS99}
J. J. Brey, J. W. Dufty, and A. Santos, J. Stat. Phys. \textbf{97},
281 (1999).








\bibitem{GK96}
A. N. Gorban and I. V. Karlin, Phys. Rev. Lett. \textbf{77}, 282
(1996); I. V. Karlin, G. Dukek, and T. F. Nonnenmacher, Phys. Rev. E
\textbf{55}, 1573 (1997); A. Santos,  Phys. Rev. E \textbf{62}, 6597
(2000).

\bibitem{SD06}
A. Santos and J. W. Dufty, Phys. Rev. Lett. \textbf{97}, 058001
(2006).



\end{thebibliography}
\end{document}